\documentclass[12pt]{iopart}

\usepackage{graphicx}
\usepackage{amsfonts}
\usepackage{color}

\begin{document}

\title{Cooperation percolation in spatial prisoner's dilemma game}

\author{Han-Xin Yang$^{1}$, Zhihai Rong$^{2}$ and Wen-Xu Wang$^{3}$}
\address{$^{1}$Department of Physics, Fuzhou University, Fuzhou
350108, China}
\address{$^{2}$Web Sciences Center, School of Computer Science and Engineering,
University of Electronic Science and Technology of China, Chengdu
610054, China}
\address{$^{3}$School of Systems Science, Beijing Normal University, Beijing 100875, China
}

\begin{abstract}

The paradox of cooperation among selfish individuals still puzzles
scientific communities. Although a large amount of evidence has
demonstrated that cooperator clusters in spatial games are effective
to protect cooperators against the invasion of defectors, we
continue to lack the condition for the formation of a giant
cooperator cluster that assures the prevalence of cooperation in a
system. Here, we study the dynamical organization of cooperator
clusters in spatial prisoner's dilemma game to offer the condition
for the dominance of cooperation, finding that a phase transition
characterized by the emergence of a large spanning cooperator
cluster occurs when the initial fraction of cooperators exceeds a
certain threshold. Interestingly, the phase transition belongs to
different universality classes of percolation determined by the
temptation to defect $b$. Specifically, on square lattices,
$1<b<4/3$ leads to a phase transition pertaining to the class of
regular site percolation, whereas $3/2<b<2$ gives rise to a phase
transition subject to invasion percolation with trapping. Our
findings offer deeper understanding of the cooperative behaviors in
nature and society.
\end{abstract}



 \maketitle
\tableofcontents
\section{Introduction} \label{sec:intro}

Cooperation is ubiquitous in biological and social systems~\cite{1}.
Understanding the emergence and persistence of cooperation among
selfish individuals remains an outstanding problem. The development
of evolutionary game theory has offered a powerful mathematical
framework to address this problem~\cite{2}. In order to capture the
interaction pattern among greedy individuals, various models have
been introduced, among which the prisoner's dilemma game (PDG) has
been a prevailing paradigm~\cite{PDG}.

In the original PDG, two players simultaneously decide whether to
cooperate or defect. They both receive the payoff $R$ upon mutual
cooperation and the payoff $P$ upon mutual defection. If one
cooperates but the other defects, the defector gets the payoff $T$,
while the cooperator gains the payoff $S$. The payoff rank for the
PDG is $T>R>P>S$. As a result, in a single round of PDG, mutual
defection is the best strategy for both players, rendering the
social dilemma, in other words, the Nash equilibrium. For simplicity
but without loss of generality, the elements of the payoff matrix
for the PDG are often rescaled with $T=b>1$, $R=1$, and
$P=S=0$~\cite{nowak}, where $b$ denotes the temptation to defect.
Since the pioneering work of Nowak and May ~\cite{nowak},
evolutionary games have been extensively studied in structured
population, including regular
lattices~\cite{lattice1,lattice2,lattice3,lattice3.1,lattice4,lattice5,lattice6,lattice8,lattice9,lattice10,lattice11,lattice13}
and complex
networks~\cite{network1,network1.1,network2,network4,network5,network6,network7,network7.1,network8,
network9,network9.1,network9.2,network9.3,network9.4,network10,network11,network12,network12.1,network13,network14}.
To facilitate cooperation in spatial PDG, a number of mechanisms
have been proposed, such as voluntary
participation~\cite{participation}, punishment~\cite{punishment},
similarity~\cite{similarity}, inhomogeneous
activity~\cite{activity}, social
diversity~\cite{diversity,diversity1}, dynamical
linking~\cite{linking}, asymmetric interaction and replacement
graph~\cite{asymmetric},
migration~\cite{migration1,migration2,migration3,migration4},
in-group favoritism~\cite{favoritism}, interdependent
links~\cite{interdependent1,interdependent2}, and so on.

It has been known that the formation of cooperator clusters plays an
important role in promoting cooperation in spatial games
~\cite{snow}. A cooperator cluster is defined as a connected
component (subgraph) fully occupied by cooperators. Within clusters,
cooperators can assist each other and the benefits of mutual
cooperation outweigh the losses against defector. It is of paramount
importance to explore the dynamical organization of cooperator
clusters so as to understand the emergence of cooperation among
selfish individuals. Previous studies focused on the size
distribution of cooperator clusters~\cite{cluster1,cluster2}, but
pay little attention to the conditions under which a giant
cooperator cluster of the order of the system size arises. In this
paper, we attempt to address this issue from the perspective of the
percolation theory~\cite{PT1,PT2}. In recent years, the percolation
transition characterized by a large spanning cluster arising at the
critical point, has been widely found in various dynamical
processes, such as the spreading of disease~\cite{disease}, the
formation of public opinion~\cite{opinion} and cascading
failures~\cite{cascading}. In particular, percolation phenomena with
respect to optimal cooperation level in diluted system were
discovered in Ref.~\cite{wang1,wang2}. However, percolation
phenomena pertaining to the emergence of cooperator clusters have
not yet been studied in the framework of evolutionary games. Here,
we report a percolation transition behavior characterized by the
emergence of a large spanning cooperator cluster when the initial
fraction of cooperators in the system exceeds a certain threshold.
The phase transition can be classified to either regular site
percolation or invasion percolation with trapping, depending on the
temptation to defect. We substantiate our findings by systematic
numerical simulations and analysis of phase transition.

The paper is organized as follows. In Sec.~\ref{sec:methods}, we
formalize the problem by introducing the PDG on square lattices. In
Sec.~\ref{sec:main results}, we study the cooperation percolation in
terms of different values of the temptation to defect. Finally,
conclusions and discussions are presented in
Sec.~\ref{sec:discussion}.

\section{Model and Methods} \label{sec:methods}

Individuals are located on a $L\times L$ square lattice with
periodic boundary conditions. Each individual $x$ can follow one of
two strategies: cooperation or defection , described by
\begin{equation}
\beta=\frac{1}{2}+\alpha^{2}
\end{equation}
respectively. At each time step, each individual plays the PDG with
its four nearest neighbors. The accumulated payoff of individual
$x$ can be expressed as
\begin{equation}
P_{x}=\sum_{y\in \Omega_{x}}z_{x}^{T}Mz_{y},
\end{equation}
where the sum runs over the nearest neighbor set $\Omega_{x}$ of node $x$
and $M$ is the rescaled payoff matrix with
\begin{equation}
M=\left(
    \begin{array}{cc}
      1 & 0 \\
      b & 0 \\
    \end{array}
  \right).
\end{equation}

According to the setting in Ref.~\cite{fraction}, initially cooperation and defection
strategies are randomly assigned to all individuals in terms of
some probabilities: with probability $f$ a node is occupied by a cooperator
and with probability $1-f$ a node is occupied by a defector. Individuals asynchronously
update their strategies in a random sequential
order~\cite{update0,update1,update2,update3}. A randomly selected
player compares its payoff with its nearest neighbors and changes
strategy by following the one (including itself) with the highest
payoff. After a period of transient time, the system will enter a
steady state.

We focus on the formation of cooperator clusters in the steady
state. In order to study the critical behavior regarding to the
giant cooperator cluster, we employ the normalized size of the
largest cooperator cluster $s_{1}$, the susceptibility $\chi$ and
the Binder's fourth-order cumulant $U$. These quantities are defined
as follows~\cite{percolation1.1}:
\begin{equation}
s_{1}=\frac{S_{1}}{N},
\end{equation}\label{eq1}

\begin{equation}
\chi=N[\langle s_{1}^{2} \rangle - \langle s_{1} \rangle^{2}],
\end{equation}

\begin{equation}
U=1-\frac{\langle s_{1}^{4} \rangle}{3\langle s_{1}^{2}
\rangle^{2}},
\end{equation}
where $S_{1}$ the size of the largest cooperator cluster, $N=L\times
L$ is the system size and $\langle \cdot \cdot \cdot \rangle$ stands
for configurational averages. According to the standard finite-size
scaling approach~\cite{percolation1.1}, the phase transition point
can be identified by the Binder's fourth-order cumulant $U$ and
there are the following power-law relationships at the critical value:
\begin{equation}
s_{1}\sim N^{-\beta/\nu} \quad and \quad  \chi\sim N^{\gamma/\nu}.
\end{equation}
We can use the tool to identify phase transitions pertaining to
the formation of cooperator clusters and explore the class to
which the phase transition belongs.

\section{Main results} \label{sec:main results}

According to our analysis of phase transition, we can separate the
values of $b$ into four regions: $b\in (1, 4/3)$, $(4/3,
3/2)$, $(3/2, 2)$ and $(2, \infty)$. We find that in the same
region, the cooperator clusters with respect to percolation transition
keep unchanged, irrespective of the value of
$b$. In the following, we present the results in the four regions
respectively.

\subsection{The case of $1< b <4/3$}

\begin{figure}
\begin{center}
\includegraphics[width=150mm]{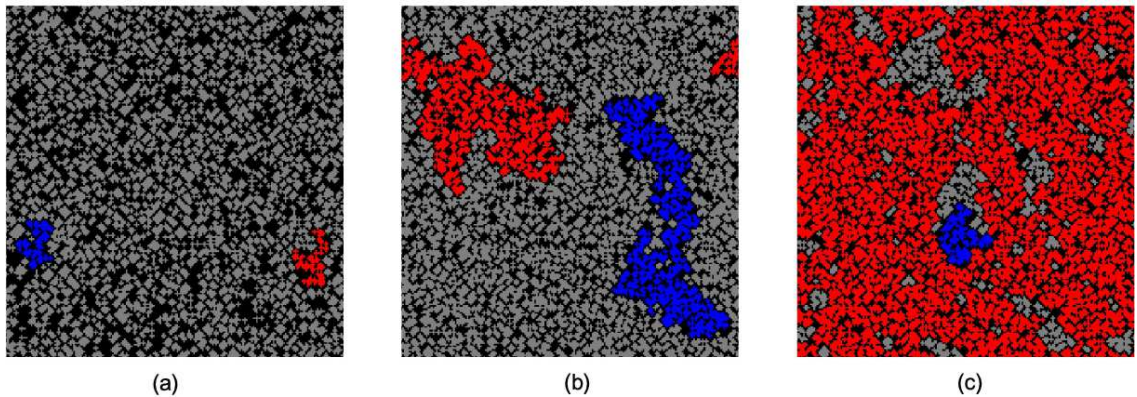}
\caption{Snapshots of distributions of cooperators and defectors on
a $200\times200$ square lattice. The temptation to defect $1< b
<4/3$. The red denotes the largest cooperator cluster, the blue
denotes the second largest cooperator cluster, the gray denotes the
other cooperator clusters and the black denotes defectors. The
initial fraction of cooperators is (a) $f=0.6$, (b) $f=0.72$ and (c)
$f=0.76$, respectively.}\label{fig1}
\end{center}
\end{figure}

We first explore the characteristics of cooperator clusters when $1<
b <4/3$. Figure~\ref{fig1} shows the snapshots of the spatial
distributions of cooperators and defectors on a $200\times200$
square lattice. We can observe that the size of the largest
cooperator cluster (denoted by red) expands as the initial fraction
of cooperators $f$ increases. In contrast, the size of the second
largest cooperator cluster (denoted by blue) becomes larger firstly
but then shrinks as $f$ continuously increases. The non-monotonic
relationship between the size of the second largest cooperator
cluster and $f$ implies the existence of a second-order phase
transition~\cite{percolation1}.

\begin{figure}
\begin{center}
\includegraphics[width=90mm]{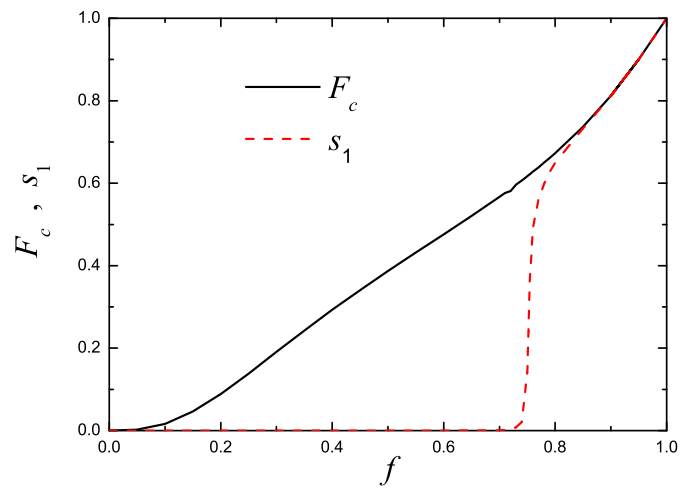}
\caption{The normalized size of the largest cooperator cluster
$s_{1}$ and the fraction of cooperators in the whole population
$F_{c}$, as a function of the initial fraction of cooperators $f$ on
a $1600\times1600$ square lattice. The temptation to defect $1< b
<4/3$. Each curve is an average of 1000 different
realizations.}\label{fig2}
\end{center}
\end{figure}

Figure~\ref{fig2} shows the normalized size of the largest
cooperator cluster $s_{1}$ and the fraction $F_{c}$ of cooperators in
the population, as a function of $f$. We see that
both $F_{c}$ and $s_{1}$ increase toward 1 as $f$ increases. Over a
wide range of $f$, $F_{c}$ is larger than $s_{1}$, where as for very
small and large values of $f$, $F_{c}$ is approximately equal to
$s_{1}$. In Fig.~\ref{fig2}, we can also find that there exists a
critical value $f_{c}$, below which $s_{1}$ approaches 0, while above
which $s_{1}$ continuously increases as $f$ increases. The fact that
$F_{c}>0.5$ at $f_{c}$ suggests that it is possible for cooperators
to form a giant cluster comparative to the size of the system,
insofar as sufficient number of cooperators survive during the
evolution.

\begin{figure}
\begin{center}
\includegraphics[width=150mm]{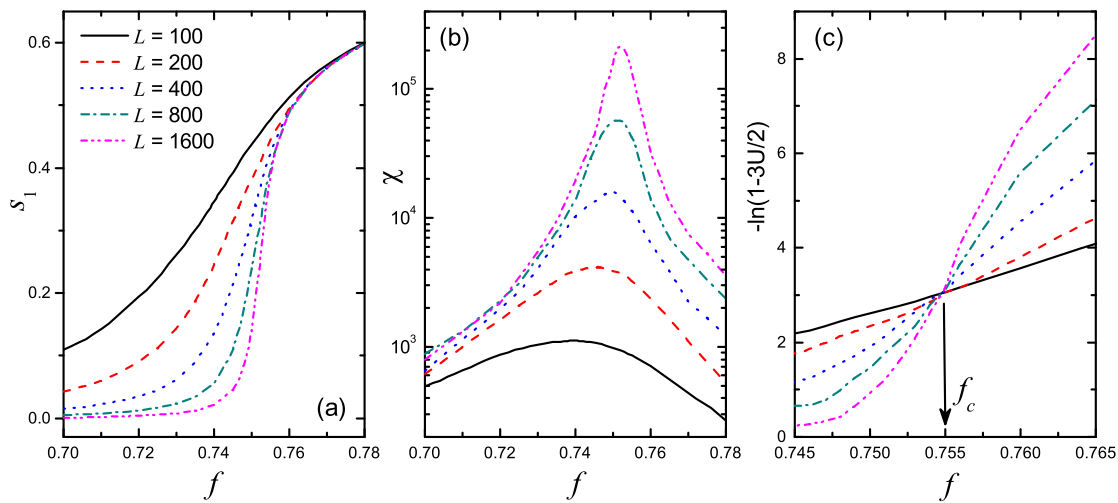}
\caption{The normalized size of the largest cooperator cluster
$s_{1}$ (a), the susceptibility $\chi$ (b) and the reduced Binder's
fourth-order cumulant $U$ (c) as a function of $f$ for different
lattice sizes $L$. In (c), all curves intersect at
$f_{c}\simeq0.7551$. The temptation to defect $1< b <4/3$. Each
curve is an average of 20000, 10000, 5000, 2000 and 1000
realizations for $L=100$, 200, 400, 800 and 1600,
respectively.}\label{fig3}
\end{center}
\end{figure}

Figure~\ref{fig3} shows the normalized size of the largest
cooperator cluster $s_{1}$, the susceptibility $\chi$ and the
reduced Binder's fourth-order cumulant $U$ as a function of $f$ for
different lattice sizes $L$. Figure~\ref{fig3}(a) shows that $s_{1}$
is almost the same for different values of $L$ for large $f$.
However, when $f$ is blow some value, $s_{1}$ decreases as $L$
increases. Figure~\ref{fig3}(b) shows that the susceptibility $\chi$
reaches a maximum at some value of $f$ for different values of $L$.
Moreover, one can see that the value of $f$ that corresponds to the
peak of $\chi$ increases with $L$. The critical value $f_{c}$ can
then be identified in Fig.~\ref{fig3}(c), where the curves of the
reduced forth-order cumulant $U$ for different $L$ intersect with
each other~\cite{BJK1,BJK2}. The intersection point gives
$f_{c}\simeq0.7551$ for $1< b <4/3$.

\begin{figure}
\begin{center}
\includegraphics[width=120mm]{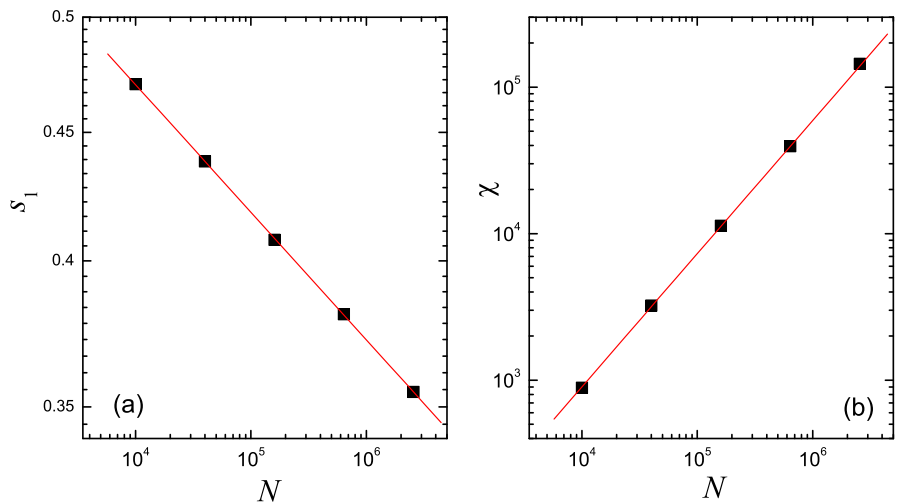}
\caption{The normalized size of the largest cooperator cluster
$s_{1}$ (a) and the susceptibility $\chi$ (b) as a function of the
system size $N$ at the critical value $f_{c}\simeq0.7551$. The
temptation to defect $1< b <4/3$.}\label{fig4}
\end{center}
\end{figure}

Figure~\ref{fig4} shows $s_{1}$ and $\chi$ as a function of $N$ at
the critical value $f_{c}$. From Figs.~\ref{fig4}(a) and (b), we
obtain the critical exponents $\beta/\nu\simeq 0.051$ and
$\gamma/\nu\simeq0.908$, which are very close to the critical
exponents of regular site percolation ($\beta/\nu\simeq 0.052$ and
$\gamma/\nu\simeq0.896$)~\cite{percolation1}. These results indicate
that the cooperation percolation belongs to the same universality
class as the regular site percolation when $1< b <4/3$.

\subsection{The case of $3/2< b <2$}

\begin{figure}
\begin{center}
\includegraphics[width=150mm]{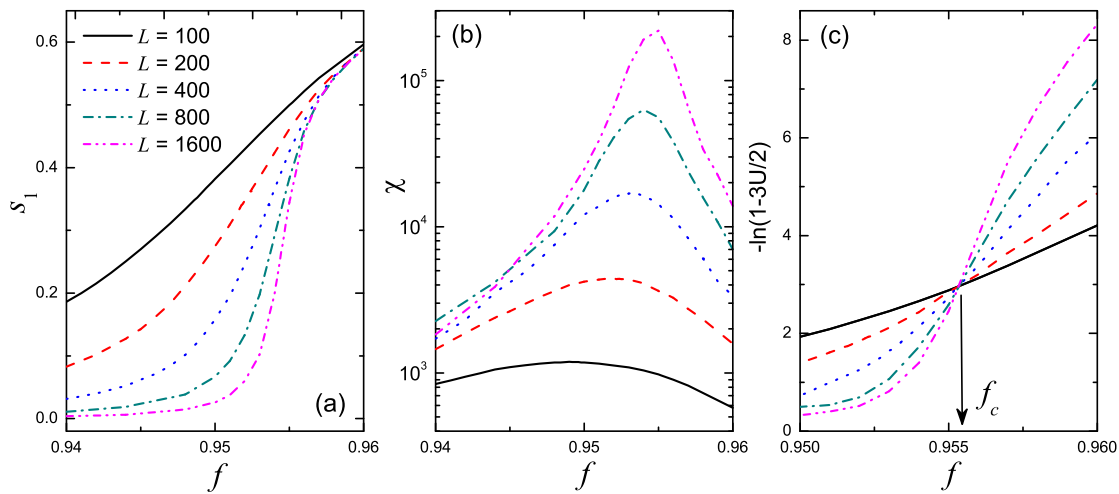}
\caption{The normalized size of the largest cooperator cluster
$s_{1}$ (a), the susceptibility $\chi$ (b) and the reduced Binder's
fourth-order cumulant $U$ (c) as a function of $f$ for different
values of the lattice size $L$. In (c), all curves intersect at
$f_{c}\simeq0.9553$. The temptation to defect $3/2< b <2$. Each
curve is an average of 20000, 10000, 5000, 2000 and 1000
realizations for $L=100$, 200, 400, 800 and 1600, respectively.
}\label{fig5}
\end{center}
\end{figure}

We explore the cooperation percolation when $3/2< b <2$.
Figure~\ref{fig5} shows the normalized size of the largest
cooperator cluster $s_{1}$, the susceptibility $\chi$ and the
reduced Binder's fourth-order cumulant $U$ as a function of $f$ for
different lattice sizes $L$. As shown in Fig.~\ref{fig5}(a), $s_{1}$
decreases as the system size increases when $f$ is below some value.
The susceptibility $\chi$ reaches the maximal value at some value of
$f$ [see Fig.~\ref{fig5}(b)] and the reduced fourth-order cumulants
cross at the critical point $f_{c}\simeq0.9553$ [see
Fig.~\ref{fig5}(c)].

\begin{figure}
\begin{center}
\includegraphics[width=120mm]{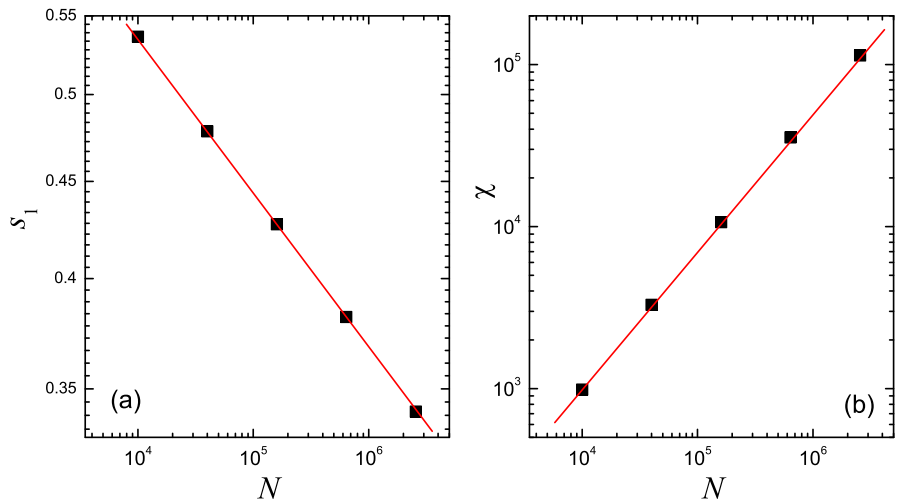}
\caption{The normalized size of the largest cooperator cluster
$s_{1}$ (a) and the susceptibility $\chi$ (b) as a function of the
system size $N$ at the critical value $f_{c}\simeq0.9553$. The
temptation to defect $3/2< b <2$.}\label{fig6}
\end{center}
\end{figure}

Figure~\ref{fig6} shows $s_{1}$ and $\chi$ as a function of the
system size $N$ at the critical value $f_{c}$. In
Figs.~\ref{fig6}(a) and (b), we obtain the critical exponents
$\beta/\nu\simeq 0.082$ and $\gamma/\nu\simeq0.856$, which are close
to the critical exponents of invasion percolation with trapping
($\beta/\nu\simeq 0.084$ and
$\gamma/\nu\simeq0.832$)~\cite{percolation2}. These results
demonstrate that the cooperation percolation belongs to the same
universality class as invasion percolation with trapping in the
region of $3/2< b <2$.

\subsection{The cases of $4/3< b <3/2$ and $b>2$}

\begin{figure}
\begin{center}
\includegraphics[width=120mm]{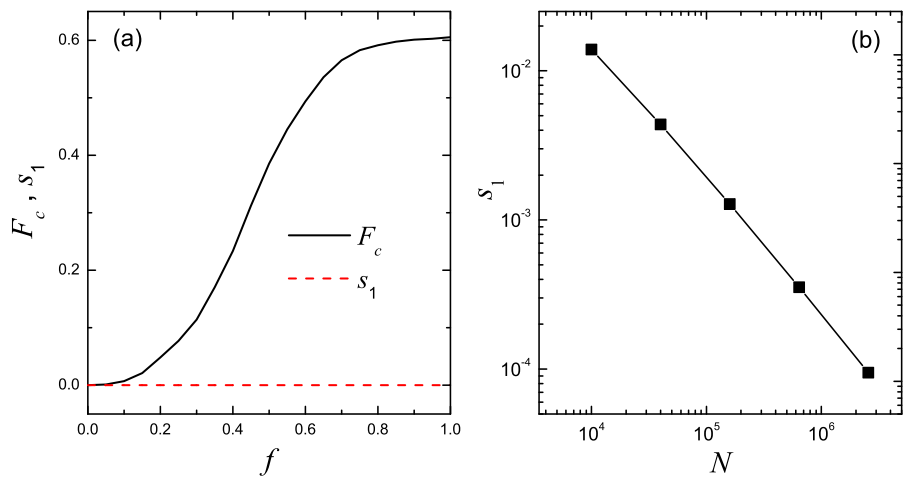}
\caption{(a) The normalized size of the largest cooperator cluster
$s_{1}$ and the fraction of cooperators in the whole population
$F_{c}$, as a function of the initial fraction of cooperators $f$ on
a $1600\times1600$ square lattice. (b) The normalized size of the
largest cooperator cluster $s_{1}$ as a function of the system size
$N$ when initially there is only one defector. The temptation to
defect $4/3< b <3/2$. }\label{fig7}
\end{center}
\end{figure}

\begin{figure}
\begin{center}
\includegraphics[width=150mm]{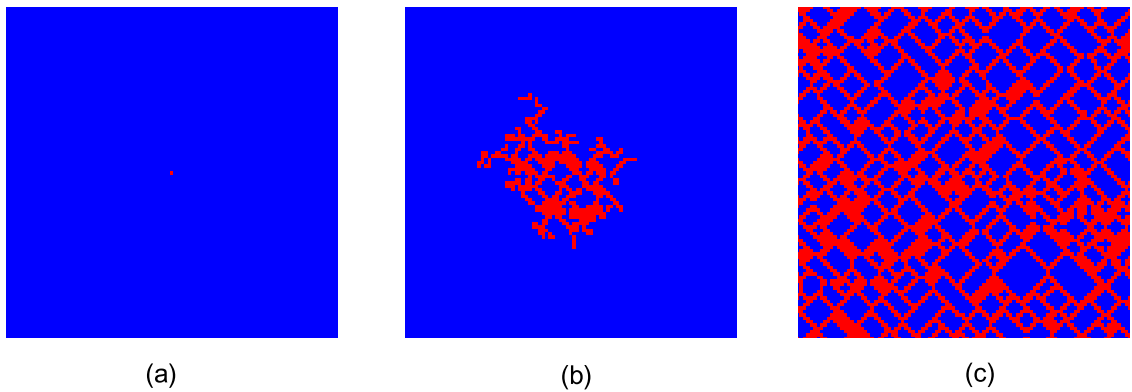}
\caption{Snapshots of distributions of cooperators (blue) and
defectors (red) at different time steps $t$ on a $99\times99$ square
lattice. The temptation to defect $4/3< b <3/2$. (a) $t=0$, (b)
$t=20$ and (c) $t=1000$. }\label{fig8}
\end{center}
\end{figure}

Figure~\ref{fig7}(a) shows the normalized size of the largest
cooperator cluster $s_{1}$ and the fraction of cooperators $F_{c}$,
as a function of the initial concentration of cooperators $f$ when
$4/3< b <3/2$. One can observe that $F_{c}$ is much lower than 1 and
$s_{1}\approx 0$ even when $f$ is very close to 1.
Figure~\ref{fig7}(b) shows $s_{1}$ as a function of $N$ when
initially there is only one defector in the system. We can find that
$s_{1}$ decreases as $N$ increases. Figure~\ref{fig7} demonstrates
that the percolation threshold for $4/3< b <3/2$ is $f_{c}=1$. To
intuitively understand why the threshold is 1 in the parameter
region, we explore the evolution of the spatial distributions of
cooperators and defectors on a square lattice with given a single
defector at the center initially ($t=0$). Figure~\ref{fig8} shows
that eventually a giant reticular defector cluster is formed and
many small cooperator clusters are separated by defectors. This
result is in agreement with that reported in
Ref.~\cite{spatialchaos}, i.e., defectors can spread over the system
even initially there is only one defector in the region of $4/3< b
<3/2$. Hence only the absence of defector can lead to a large
cooperator cluster that covers the whole lattice, accounting for the
phase transition at $f_{c}=1$.

In the case of $b>2$, cooperators cannot survive when $f<1$ (results
are not shown here), implying that even one defector can lead to the
extinction of cooperators in the ocean of cooperators when $b>2$. We
can thus infer that the percolation threshold for $b>2$ is also
$f_{c}=1$.

\section{Conclusions and Discussions} \label{sec:discussion}

In conclusion, we have explored the formation of cooperator clusters
in the prisoner's dilemma game, finding that the process of
establishing a giant cluster pertains to the percolation behavior.
In particular, when the initial fraction of cooperators in the
system exceeds a critical threshold, there arises a giant spanning
cooperator cluster, resulting from the merging of many small
cooperator clusters. The phase transition behavior is validated by
various scaling laws at the critical point, such as the normalized
size of the largest cooperator cluster and the susceptibility scale
with the system size. Strikingly, the phase transition belongs to
different universality classes, depending on the temptation of
defect. The results on square lattices demonstrate that the phase
transition is subject to the class of regular site percolation when
$1< b <4/3$ or invasion percolation with trapping when $3/2< b <2$.
Whereas in the parameter region $4/3< b <3/2$ and $b>2$, the
percolation threshold is 1, indicating that even one defector can
prevent the formation of large cooperator clusters. Interestingly,
we found that the partition of the parameter region in terms of
percolation is exactly the same as previous findings based on the
chaotic spatial patterns in literature~\cite{nowak,spatialchaos}.
The agreement offers an underlying connection between the phase
transition and the spatial chaos in evolutionary games.

Our findings presented here raise a number of questions, answers to
which could further deepen our understanding of the persistence and
dominance of cooperation in terms of a large cooperator cluster. For
example, for strategy updating rules rather than the currently used
best-take-over rule, such as the Fermi rule~\cite{update0}, if the
percolation transition remains? If the answer is positive, will the
universality class change? Another significant question pertaining
to the network structure is how does the structural property affect
the percolation phenomena, e.g., small-world and scale-free
topology. Taken together, our results indicate that cooperation in
many evolutionary games can be explored from the perspective of
cooperator clusters in the combination with the tools for
quantifying phase transition and percolation, opening new avenues to
deepening our understanding of cooperative behaviors widely observed
in many aspects in nature and society.

\section*{Acknowledge}
This work was supported by the National Natural Science Foundation
of China (Grants No. 11247266, No. 11105011 and No. 61004098), the Natural
Science Foundation of Fujian Province of China (Grant No.
2013J05007), and the Research Foundation of Fuzhou University (Grant
No. 0110-600607).

\end{document}